\documentclass[twocolumn,showpacs,preprintnumbers,amsmath,amssymb]{revtex4} 

\usepackage{graphicx}
\usepackage{dcolumn}
\usepackage{bm}
\usepackage{psfrag}
\usepackage{subfigure}

\begin{document}


\title{Curvature Constraints from the Causal Entropic Principle\\}    

\author{Brandon Bozek, Andreas Albrecht, and Daniel Phillips}                             
\affiliation{Physics Department, University of California, Davis.}                                                                                                                                                                                                                                               
\date{\today}                                                                                               

\begin{abstract}

Current cosmological observations indicate a preference for a
cosmological constant that is drastically smaller than what can be
explained by conventional particle physics. The Causal Entropic Principle (Bousso,
{\it et al}.) provides an alternative approach to anthropic attempts
to predict our observed value of the cosmological constant by
calculating the entropy created within a causal diamond. We have
extended this work to use the Causal Entropic Principle to predict the
preferred curvature within the "multiverse". We have found that values larger
than $\rho_k = 40\rho_m$ are disfavored by more than $99.99\%$ and a peak
value at $\rho_{\Lambda} = 7.9 \times 10^{-123}$ and $\rho_k =4.3
\rho_m$ for open universes. For universes that allow only positive
curvature or both positive and negative curvature, we find a
correlation between curvature and dark energy that leads to an
extended region of preferred values. Our universe is found to be disfavored to an extent
depending the priors on curvature. We also provide a
comparison to previous anthropic constraints on open universes and
discuss future directions for this work. 
\end{abstract}

\maketitle

\section{\label{sec:level1} Introduction}

The simplest explanation for the observed acceleration of the universe
is Einstein's cosmological constant, $\Lambda$. However, the value
that explains the acceleration is many orders of magnitude smaller
than that expected from quantum field theory. We are then left either to
determine a method to set the cosmological constant to a small
value or to consider $\Lambda$ an environmental variable varying from place to place in the multiverse.

Following the environmental approach numerous authors (many inspired
by the pioneering work of Weinberg\cite{Weinberg:1987dv}), have sought to explain the observed value of $\Lambda$ by postulating that the most likely universe to be observed would be
that which contains the largest potential to contain
observers. However, such ``anthropic'' approaches can become burdened by
complicated assumptions on the nature of observers. In their ``Causal
Entropic Principle'' (CEP) Bousso { \it et al.} \cite{Bousso:2007kq}
took this reasoning in a simple and elegant direction by associating
observers with entropy increase. Initial applications of this approach
have successfully predicted our value of
$\Lambda$\cite{Bousso:2007kq,Cline:2007su}.  The CEP has added  
appeal because there has been long standing (if not
universal\cite{FeynmanMess}) acceptance of the idea that entropy increase
would need to be imposed as a condition specific to observers rather
than a global and eternal property of the Universe\cite{Boltzmann:1895xx,Albrecht:2002uz,Dyson:2002pf,Albrecht:2004ke}. 
Specifically, the CEP gives a weight to each set of cosmological
parameters proportional to the entropy produced within a causal
diamond in the corresponding cosmology.  In addition to the original
work\cite{Bousso:2007kq} which found our value of $\Lambda$ to be
within one sigma of the peak of their predicted probability
distribution, the CEP was further developed by  Cline {\it et
al.} \cite{Cline:2007su}, exploring constraints on other cosmological
values such as density contrast, baryon fraction, matter abundance,
and dark matter annihilation rate.  

In this paper we develop this method further by using CEP to
jointly predict the values of curvature and $\Lambda$ most likely to
be observed. Since Cline {\it et al.}
\cite{Cline:2007su} did not find 
significant features in their extended parameter space, we chose to vary only
$\rho_k$ and $\rho_{\Lambda}$ for this work and hold all other
parameters fixed. However, given the tail in the probability
distribution for positive curvature, an interesting extension to this
work would be to also vary additional parameters such as the density
contrast.

Anthropic constraints on curvature are interesting in their own right,
in the context of the ``flatness
problem''\cite{Dicke:1979ig,Guth:1980zm} which suggests that in the
absence of something like cosmic
inflation\cite{Guth:1980zm,Linde:1981mu,Albrecht:1982wi} the ``most
natural'' realization of big bang cosmology would be highly dominated
by curvature.  A number of authors have already considered anthropic
bounds on
curvature\cite{Vilenkin:1996ar,Garriga:1998px,Freivogel:2005vv}, but 
we believe this is the first work to apply the CEP to curvature. We
find that our results place an upper limit on the allowed negative curvature as
expected, but one that is looser than those works mentioned above due to a more lenient tolerance on
the sizes of structure that are allowed to form. Further, we find that
our peak probability for open universes to be away from the upper edge of our probability distribution, not dictated by it, as would be the case in the other work. We also consider solutions that allow for just
positive curvature and for both positive and negative curvature. In
these cases we find a tail in the probability distribution that allows
for a wide range of allowed $\Lambda$ and $\rho_k$, a large fraction
of which are significantly larger than our measured values. We find
that our universe is  not ruled out in any scenario, but, depending on
one's choice of priors on curvature, disfavored to a certain
degree.

In section \ref{sec:Sec2} we review the CEP and the cosmology we will be considering. We then review the star formation model we will use in section \ref{sec:Sec2b}. We discuss our results in section \ref{sec:Sec3} and our conclusions in section \ref{sec:Sec4}.

\section{\label{sec:Sec2} The Causal Entropic Principle}
 
 \begin{figure*}[!t]
\begin{center}
\includegraphics[width=0.32\textwidth]{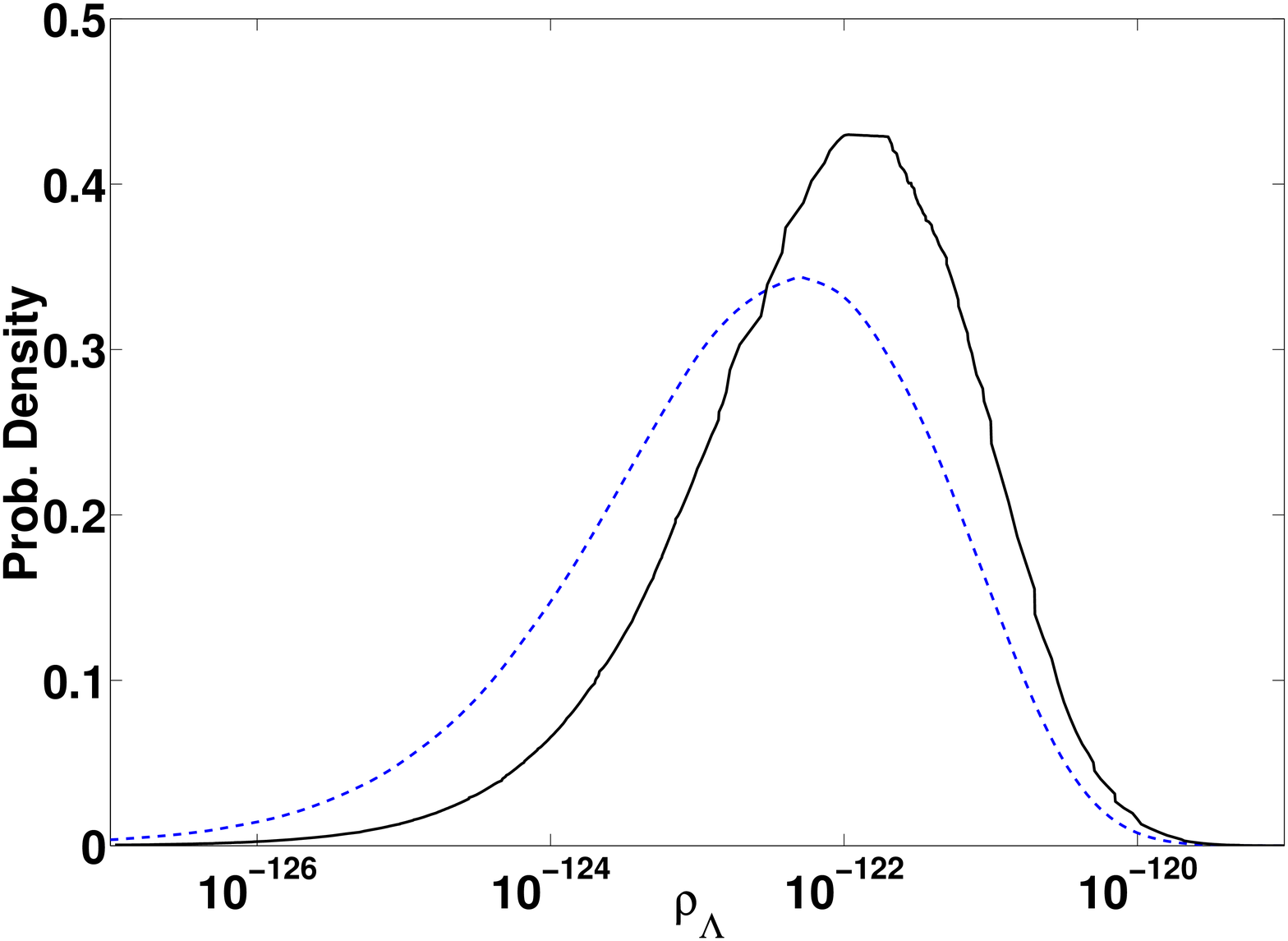}
\includegraphics[width=0.32\textwidth]{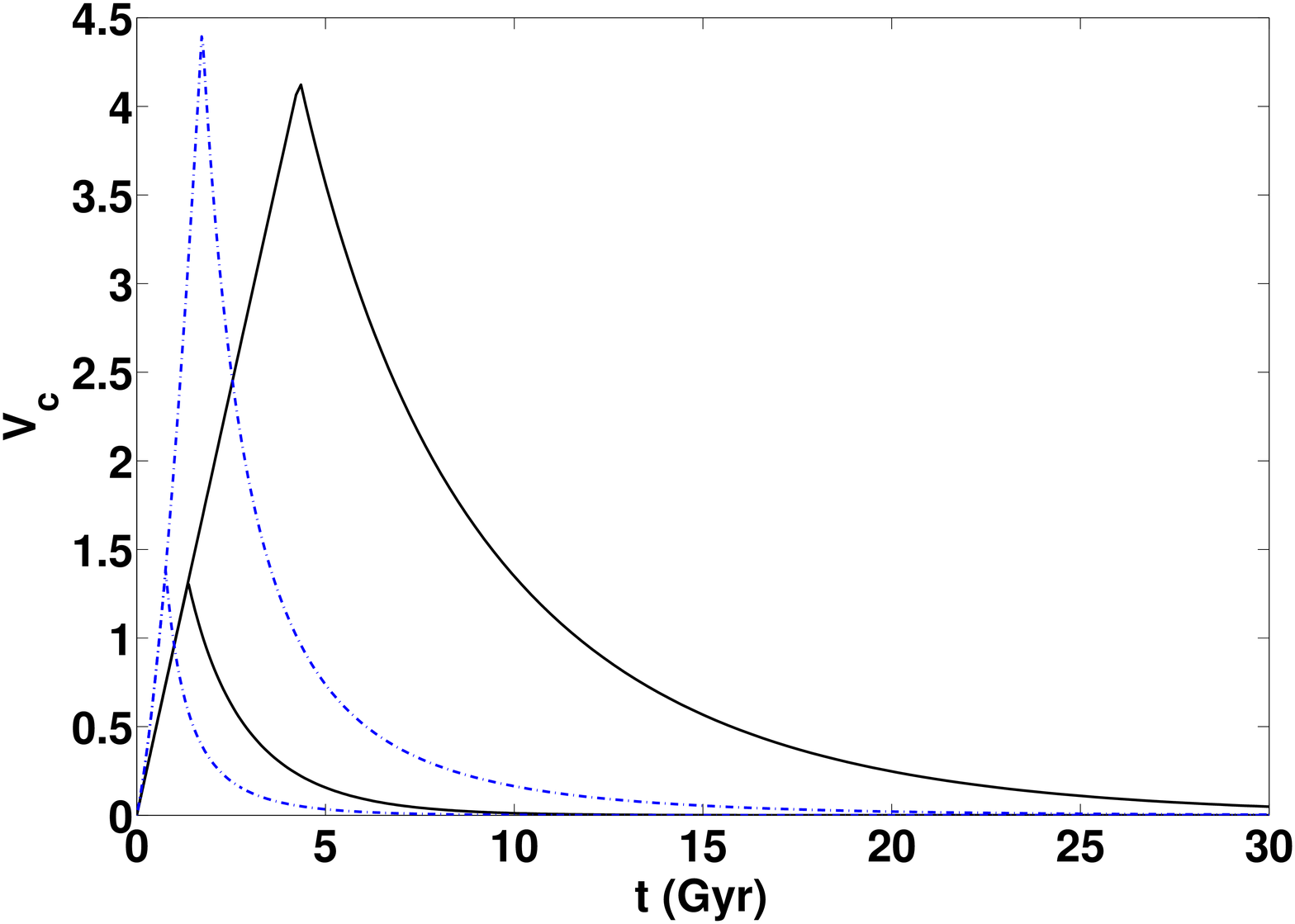}
\includegraphics[width=0.32\textwidth]{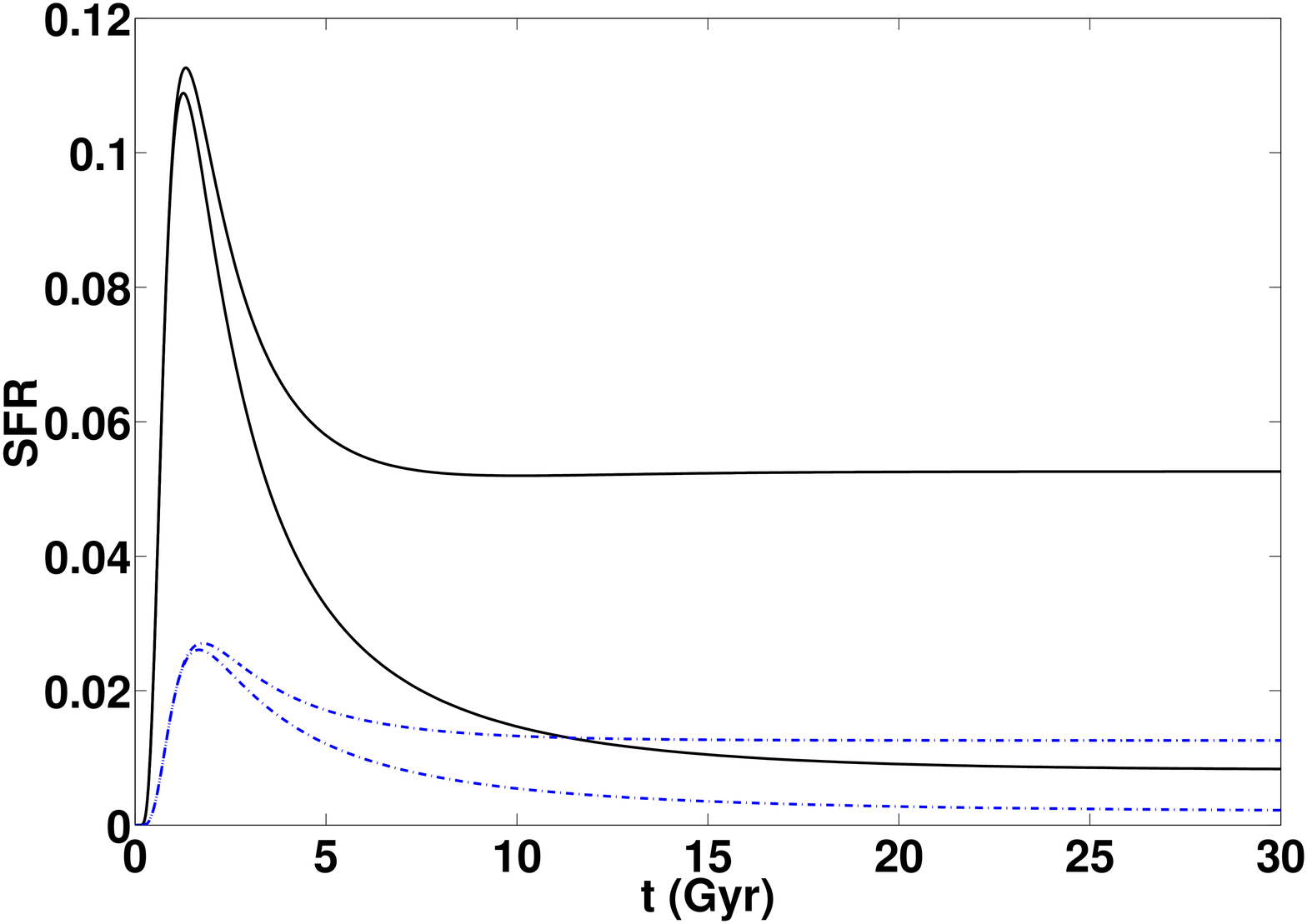}
\includegraphics[width=0.32\textwidth]{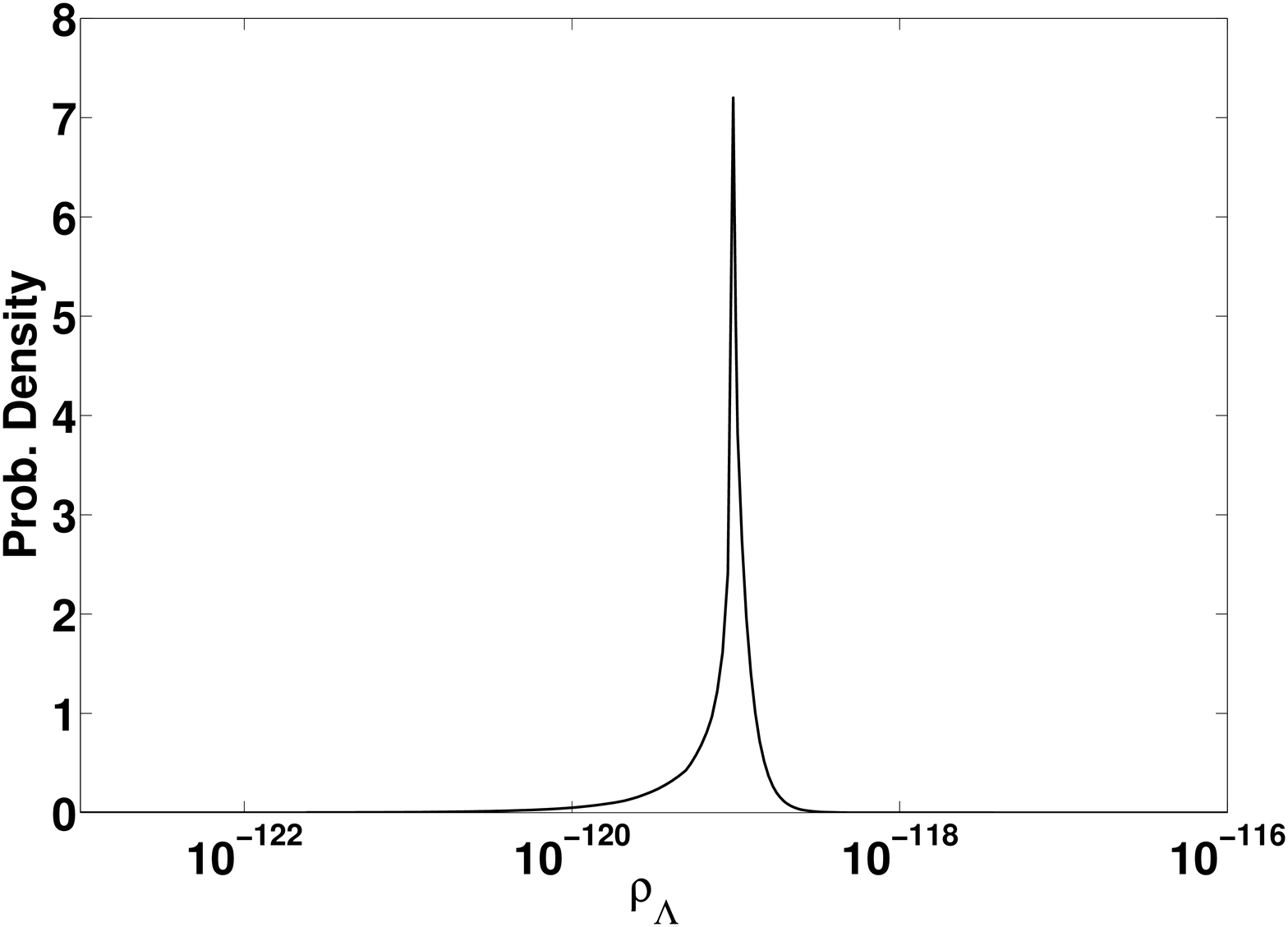}
\includegraphics[width=0.32\textwidth]{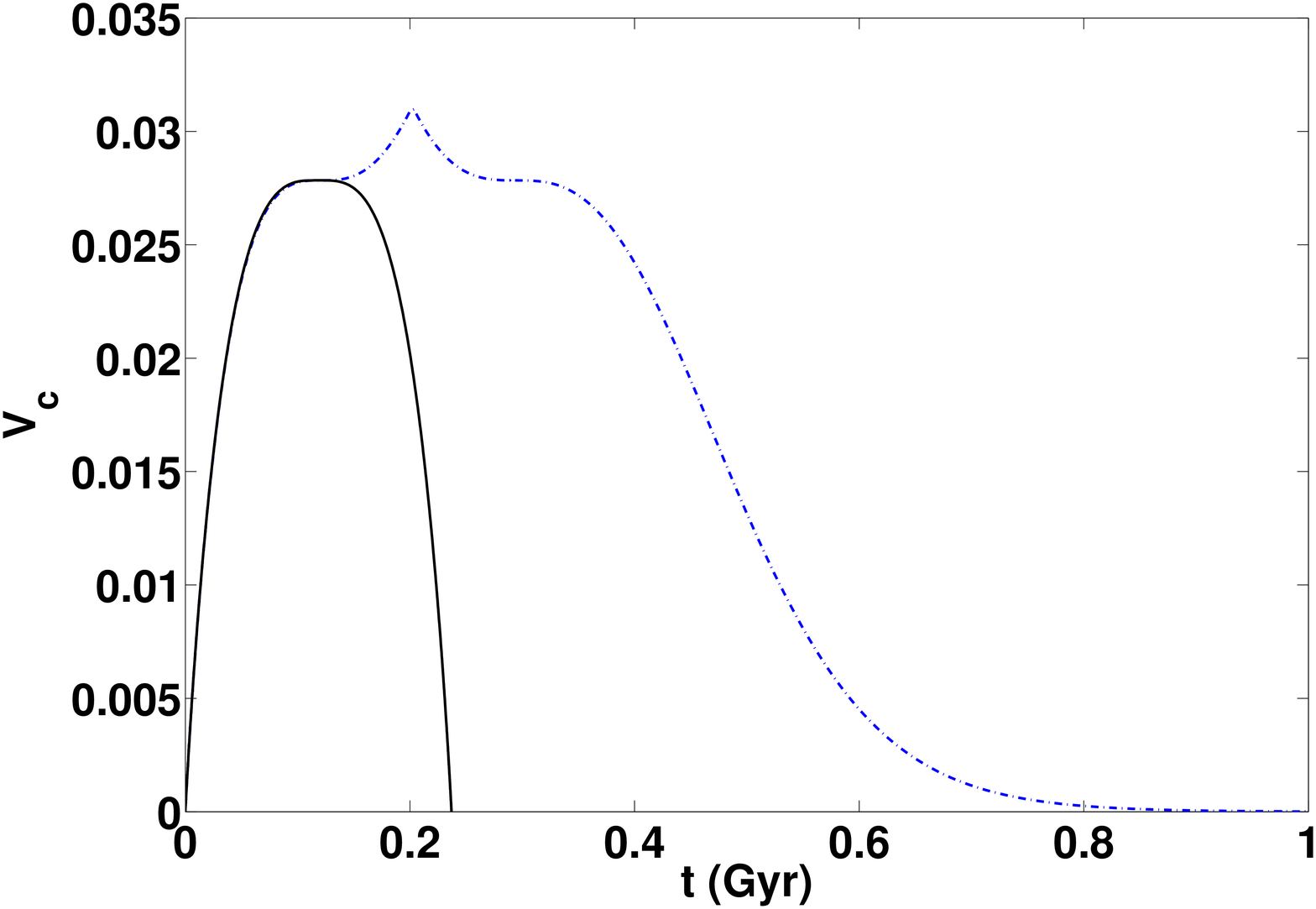}
\includegraphics[width=0.32\textwidth]{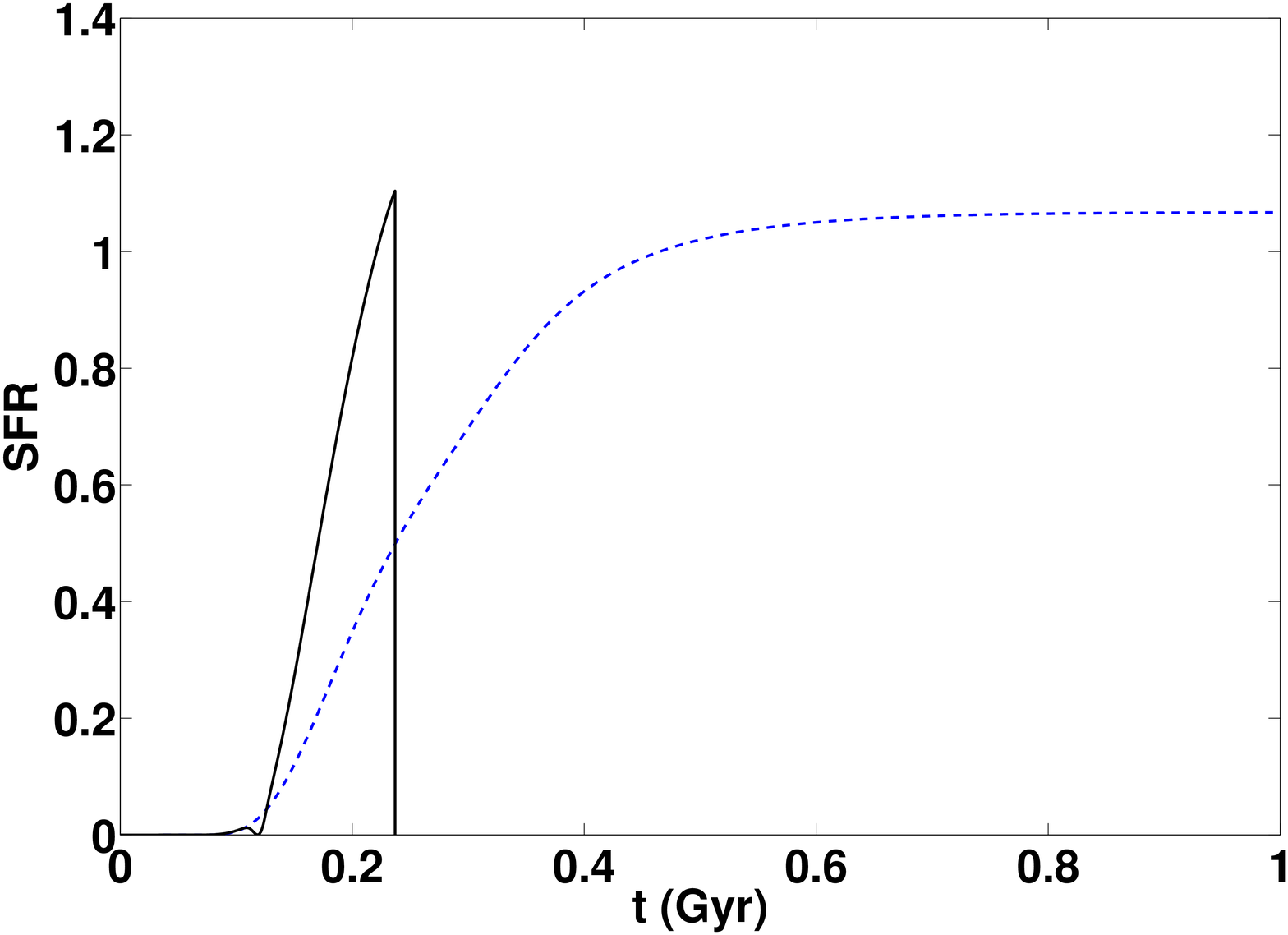}
\caption{Top Left Panel: Probability density for $\rho_\Lambda$ with
  fixed curvature of $\rho_k = 10\rho_m$ (dashed) and $\rho_k =
  0$ (solid). Top Center Panel:  The comoving
  volume (in units of $10^{12} Mpc^3$) for $\rho_{\Lambda} =
  10^{-123}$ (larger) and $\rho_{\Lambda} = 10^{-122}$ (smaller) are
  shown for $\rho_k = 10\rho_m$ (dashed) and $\rho_k = 0$
  (solid). Top Right Panel: The star formation rate in units of
  $\frac{M_{\odot}}{Mpc^3 yr}$. The upper curve for
  each value of curvature represented in solid and dashed respectively is
  $\rho_{\Lambda} = 10^{-122}$ and the lower is $\rho_{\Lambda} =
  10^{-123}$. The bottom row is the probability density, comoving
  volume, and star formation rate for $\rho_k = -50\rho_m$. The
  blue/dashed curve is $\rho_{\Lambda} = 10^{-119}$ and the
  black/solid curve is  $\rho_{\Lambda} = 10^{-120}$. 
} 
\label{fig:rhoandprob}
\end{center}
\end{figure*}
 
	According to the CEP the probability distribution for $\Lambda$ is given by the equation:	
\begin{equation} 
\label{eq: prob}
\frac{d^2P}{d\rho_{\Lambda} d\rho_k} = P_0 \times w(\rho_{\Lambda},\rho_k) \times \frac{d^2p}{d\rho_{\Lambda} d\rho_k}
\end{equation}	
where $w(\rho_{\Lambda},\rho_k)$ is a weighting factor, $P$ is the total
probability, $P_0$ is a normalization factor, and $p$ is the total
prior. We will assume the joint prior probability of
$p(\rho_{\Lambda},\rho_k)$ to be independent giving
$p(\rho_{\Lambda},\rho_k) = p(\rho_{\Lambda}) \times
p(\rho_k)$. The prior for $\Lambda$ is an expression of how the
``multiverse'' is populated by physics with different values of
$\Lambda$. Here we use the standard form (sometimes motivated
by the string theory landscape) taken by previous
authors\cite{Weinberg:1987dv,Bousso:2007kq}: $\frac{dp}{d\rho_{\Lambda}} = {\rm constant}$.  For simplicity we also take
$\frac{dp}{d\rho_k} = {\rm constant}$, which will enable a discussion of
the flatness problem later in the paper.  These ``flat'' priors mean the
largest allowed ``cutoff'' values of $\rho_k$ and $\rho_\Lambda$ set the typical 
values for the prior. The value of the cutoff turns out to be
unimportant because for flat priors $w(\rho_\Lambda,\rho_k)$ dictates
the shape of the final probability 
distribution. 

In the CEP framework we set $w(\rho_{\Lambda},\rho_k) = \Delta S$, where
$\Delta S$ is the total entropy produced within a causal
diamond. After considering numerous astrophysical sources for entropy
production, Bousso {\it et al.} \cite{Bousso:2007kq} find that the dominant
form of entropy production is star light reradiated by
dust. As in \cite{Bousso:2007kq}, $\Delta S $ is given by.
\begin{equation}
\label{eq: totentopy}
\Delta S = \int_{t_i}^\infty \frac{d^2S}{dV_c dt} V_c dt
\end{equation}
where $V_c$ is the total comoving volume of an observer's causal
patch. A causal patch is defined by a future light cone taken at an
initial point, such as reheating following inflation, intersected by a
past light cone at a late time point, which in the case of a universe dominated by a
cosmological constant is bounded by a de Sitter horizon and in the case of a universe dominated by positive curvature the late time event is the crunch.

We use the metric:
\begin{equation}
ds^2 = -dt^2 + a(t)^2 R_0^2 [d\chi^2 + S_k(\chi)^2d\Omega^2]
\end{equation}
where $S_k(\chi) = \sin(\chi)$ for positive curvature, $S_k(\chi) = \sinh(\chi)$ for negative curvature, and $S_k(\chi) = \chi$ for no curvature. The causal diamond is then given by $R_0 \chi = \frac{ \Delta \tau}{2} - |\frac{ \Delta \tau}{2}  + \tau|$, where $\tau = \int \frac{dt}{a(t)}$. The comoving volume is then:
\begin{equation}
V_c = \left\{ 
\begin{array}{l l}
  2 \pi R_0^3 [\chi -\frac{1}{2}\sin(2 \chi)] & \quad \mbox{for k = +1}\\
   \frac{4\pi}{3} R_0^3 \chi^3 & \quad \mbox{for k = 0}\\
  2 \pi R_0^3 [\frac{1}{2}\sinh(2 \chi) - \chi] & \quad \mbox{for k = -1}\\

\end{array} \right. 
\end{equation}

The scale factor $a(t)$ can be found by solving the Friedmann equation:  
\begin{equation}
H^2 = \frac{8 \pi}{3} (\frac{\rho_m}{a^3} + \rho_{\Lambda} + \frac{\rho_k}{a^2})
\end{equation}
where $\rho_{\Lambda} = {\Lambda}/{8 \pi}$, $\rho_k = {-3
  k}/{8 \pi R_0^2}$, and $k = \{-1,0,1\}$ for a negative, flat, and
  positively curved universe respectively.  The value of the matter
  density today 
  ($a=1$) is set at $\rho_m = 5.2\times10^{-124}$ in Planck units which we use throughout
  unless otherwise noted. Following previous work in the topic we neglect radiation. In this work we hold $\rho_m$ fixed and allow the curvature today, $\rho_k$, and $\rho_\Lambda$ to vary.

The other part of Eqn. \ref{eq: totentopy}, $\frac{d^2S}{dV_c dt}$, is the
entropy produced per comoving volume per time, which is calculated by
the convolution
\begin{equation}
\frac{d^2S}{dV_c dt}(t)= \int_0^{t} \frac{d^2S}{dMdt'}(t -t') \dot{\rho_{\star}}(t')dt'
\end{equation}
where $\frac{d^2S}{dMdt'}(t -t')$ is the entropy production rate per stellar mass
at time $t$ due to stars born at an earlier time, $t'$, and $\dot{\rho_{\star}}(t')$ is the
star formation rate at $t'$. The entropy rate per stellar mass is found by calculating 
\begin{equation} \label{eq:stellarmass}
\frac{d^2S}{dMdt'}(t -t') = \frac{1}{\left<M\right>}\int_{0.08M_{\odot}}^{M_{max}(t -t')} \frac{d^2s}{dN_{\star}dt} \xi_{IMF}(M) dM
\end{equation}
where $\xi_{IMF}(M)$ is the initial mass function and
$\frac{d^2s}{dN_{\star}dt}$ is the entropy production rate for a
single star. The latter is given by the stellar luminosity divided by
the effective temperature. The number of photons emitted by a star is
dominated by the half that are reprocessed by dust at an effective
temperature of $20 \text{mev}$. This is given by: 
\begin{equation}
\frac{d^2s}{dN_{\star}dt} = \frac{L_{\star}}{T_{eff}} = \frac{1}{2} (\frac{M}{M_{\odot}})^{3.5} \, 3.7\times10^{54} yr^{-1}. 
\end{equation}
The prefactor in Eqn. \ref{eq:stellarmass} is the average initial mass,
$\left<M\right> = 0.48 M_{\odot}$. The lower limit of Eqn. \ref{eq:stellarmass} is
the minimum mass of a star that can support nuclear burning.
Following Bousso, we take the upper limit to be
\begin{equation}
M_{max}(t -t') = \left\{ 
\begin{array}{l l}
  100M_{\odot} & \quad \mbox{for $t - t' < 10^5$ yr}\\
  (\frac{10^{10} yr}{t-t'})^{0.4} M_{\odot} & \quad \mbox{for $t - t' > 10^5$ yr}\\
\end{array} \right. 
\end{equation}

\section{\label{sec:Sec2b} Star Formation}
A key aspect to this work is how well the star formation rate is
modeled. While Bousso {\it et al.} \cite{Bousso:2007kq}  use star formation
rates of Nagamine \cite{Nagamine:2006kz} {\it et al.} and Hopkins and Beacom
\cite{Hopkins:2006bw}  with some simple modifications to extend these
models to include different cosmological constant values, we will
follow Cline {\it et al.} \cite{Cline:2007su} who use a
model proposed by Hernquist and Springel (HS)
\cite{Hernquist:2002rg}. The HS model was found to produce similar
results to those found in 
Bousso {\it et al.} but was more straightforward to extrapolate to the
case were a larger number cosmological parameters are varied.  The HS
star formation model is given by this equation: 
\begin{equation}
\label{eq: SFR}
\dot{\rho_{\star}} = \rho_{m} s_0 q(t) (1 - \text{erf}(\sqrt{\frac{a}{2}} \frac{\delta_c}{\sigma_4}))
\end{equation}
where $\delta_c = 1.6868$, $a = 0.707$, $s_0 = 3.7995 \times 10^{-63}$ (taken from \cite{Hernquist:2002rg}), and $\rho_m$ (same as above) are constants and $\sigma_4$ 
is the root-mean-square density fluctuation that corresponds to the
mass scale that virializes at a temperature of $T = 10^4 K$.

The star
formation efficiency, $q(t)$, encompasses the rate and efficiency of
radiative cooling within a collapsing object that leads to star
formation. HS model this process with: 
\begin{equation}
q(t) = (\frac{\chi(t) \tilde{\chi}}{(\chi(t)^m + \tilde{\chi}^m)^{\frac{1}{m}}})^p
\end{equation}
where $\chi(t) = (\frac{H}{H_0})^{2/3}$. $\tilde{\chi} = 4.6$, $m =
6$, and $p = 2.72$ are constants fit from numerical simulations and
$H_0 = 70 \frac{km/s}{Mpc}$.  
For universes with positive curvature that end
in a crunch, the star formation rate of Eqn. \ref{eq: SFR} continues up until the crunch. We therefore needed to
place a bound on late time star formation when we no longer trust
our model. We set $\dot{\rho_{\star}} = 0$ when $\rho_r =\rho_m$ in the
collapsing phase, where $\rho_r = 1.5\times10^{-127}$. This choice allows for an exploration of the CEP properties without a strong limiting effect put in by hand.

Star formation rates for several different values of curvature and
$\Lambda$ are shown in the right panel of Fig.
\ref{fig:rhoandprob}. Their corresponding probability curves are shown
in the left panel of Fig. \ref{fig:rhoandprob}. 

\section{\label{sec:Sec3} Results and Discussion}

\begin{figure}
\includegraphics[width=0.45\textwidth]{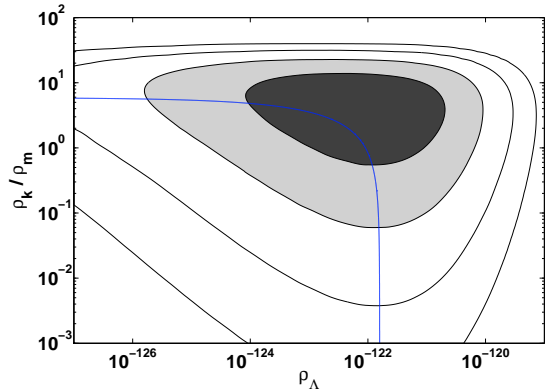}

\caption{The 68.27\% (dark grey), 95.44\% (light grey), 99.73\% (inner white), and 99.99\% (outer white) contours for a multiverse that only admits negative curvature universes. The solid blue line that cuts through the other 4 contours is the anthropic bound from \cite{Freivogel:2005vv} for smaller sized galaxies.}  
\label{fig:justnegcurv}
\end{figure}

\begin{figure}
\includegraphics[trim = 30mm 30mm 0mm 30mm, clip, width=0.57\textwidth]{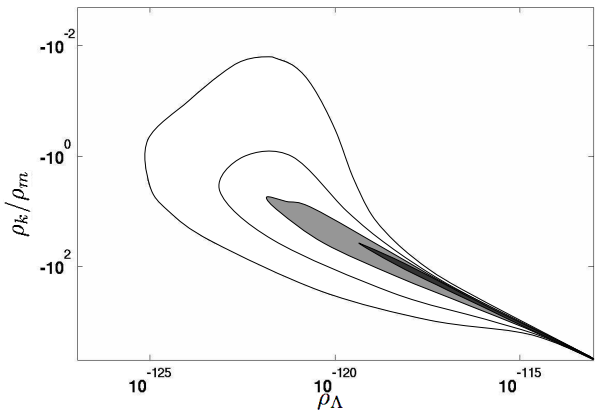}

\caption{The 68.27\% (dark grey), 95.44\% (light grey), 99.73\% (inner white), and 99.99\% (outer white) contours for a multiverse that only admits positive curvature universes.}  
\label{fig:justposcurv}
\end{figure}

\begin{figure}
\includegraphics[width=0.45\textwidth]{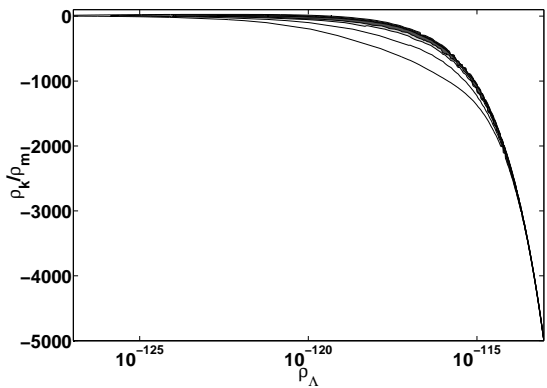}

\caption{The 68.27\% (dark grey), 95.44\% (light grey), 99.73\% (inner white), and 99.99\% (outer white) contours for a multiverse that admits both positive and negative curvature universes.}  
\label{fig:posandnegcurv}
\end{figure}

Using the equations of the previous section we solve for the
probability distribution over a wide range of open and closed
universes.  To enable the most general discussion we consider both
open and closed cosmologies together and separately. (It is commonly \cite{Freivogel:2005vv}
but not universally \cite{Buniy:2006ed} thought that the string theory
landscape only leads to the open case.)

If we were to only consider open universes, then
Fig. \ref{fig:justnegcurv} depicts the resulting probability density distribution
in $\log\rho_{\Lambda}$-$\log\rho_k$
space.  Values of curvature of $\rho_k > 40\rho_m$ fall outside of
the $99.99\%$ CL. Smaller values of $\Lambda$ and curvature lead to
larger causal diamonds and therefore have the most total entropy
production, as depicted in Fig. \ref{fig:rhoandprob}. This balances
the majority of vacua having larger values of both $\Lambda$ and
curvature, giving a peak value at $\rho_{\Lambda} = 7.9\times
10^{-123}$ and $\rho_k = 4.3 \rho_m$. Using the upper bound ($95\%$
CL) on negative curvature from WMAP+HST \cite{Komatsu:2008hk}, our
universe of $\rho_{\Lambda} = 1.25 \times 10^{-123}$ and $\rho_k =
0.016 \rho_m$ is in the $99.73\%$ CL. Fig. \ref{fig:rhoandprob}
illustrates that for a fixed value of negative curvature the
distribution for $\Lambda$ remains roughly unchanged from the
distribution for a flat universe.

Now considering only positively curved universes, there is a clear
correlation between $\Lambda$ and curvature which comes from
competition between positive curvature and $\Lambda$.
 This can cause the universe to ``loiter''\cite{Sahni:1991ks} in a
 state with little cosmic expansion but plenty of structure growth.
These conditions  conspire to create both a larger causal
diamond and enhanced linear growth.  This results in the tail on the
bottom right of Fig. \ref{fig:justposcurv} where there is ridge between collapsing
regions to the left and non-collapsing regions to the right in the
$68\%$ CL. For a fixed value of positive curvature, small values of
$\Lambda$ lead to a universe that will recollapse before there is
significant star formation. As $\Lambda$ is increased the recollapse
is delayed allowing for more star formation and a larger causal
diamond, giving more total entropy produced and therefore a more
likely universe within the CEP framework. This continues until
$\Lambda$ is large enough to allow for a non-collapsing universe, at
which point larger values of $\Lambda$ begin to suppress growth. The
narrowing of the tail comes from deviations from the ridge having
large energy densities that lead to either a rapid recollapse or early
$\Lambda$ domination. Using the upper bound ($95\%$ CL) on positive
curvature from WMAP+HST \cite{Komatsu:2008hk}, our universe of
$\rho_{\Lambda} = 1.25 \times 10^{-123}$ and $\rho_k = -0.06 \rho_m$
is in the $99.99\%$ CL. 

Fig. \ref{fig:posandnegcurv} shows the 2
dimensional probability density distribution, $\frac{d^2P}{d\log\rho_{\Lambda} d\rho_k}$,  in $\log\rho_{\Lambda}$-$\rho_k$
space for both positive and negative curvature. 
The full span range of  curvature allowed by WMAP+HST
\cite{Komatsu:2008hk} ($-0.06 \rho_m \le \rho_k \le 0.016 \rho_m$) is
in the $95.44\%$ CL.   A significant fraction of the 
values within the  $68.27\%$ CL are positively curved 
universes of both large amounts of curvature and dark energy compared
with our universe due to the competing effects mentioned above leading
to a similar tail on the lower right.

A recent paper by Bousso and Leichenauer \cite{Bousso:2008bu} has
argued that the asymptotic behavior of the star formation model shown
in upper right panel of Fig. \ref{fig:rhoandprob} may be
unphysical. Since the CEP framework depends on an accurate accounting
of star formation in universes far different than ours, a careful
study of different models is an important aspect of developing this
work further. However, we suspect that the asymptotic behavior leads
to a subdominant effect on the final probability distributions since
it coincides with a decreasing comoving volume that will diminish the
contribution it will make to the total entropy contribution.  

On the other hand, the bottom right portion of the tail in
Figs. \ref{fig:justposcurv} and \ref{fig:posandnegcurv} is an area
where we have little confidence in our star formation model as the
duration of matter domination is increasingly smaller as we move
further out onto the tip of the distribution. A different cut on late
time star formation from the one we chose above or another star
formation model may find the bottom right tip less favored or ruled
out.  

We have extended the CEP to include curvature and found that
regardless of whether one considers both positive and negative
curvature or just one of the two options, a non-zero value of
curvature appears to be preferred. We have also found that our
universe is not ruled out in any scenario considered here, but is
somewhat disfavored in some scenarios. The favored values for an open
universe are just a few orders of 
magnitude larger than values favored by modern data, so to the extent
that the flatness puzzle is about why the curvature is not given by the
Planck scale, the CEP seems to put a significant dent in the
flatness puzzle.  This is not dissimilar to anthropic arguments of
curvature, where structure formation is cut off 
by excessive curvature, however the CEP offers a less restrictive
initial assumption. In Fig. \ref{fig:justnegcurv} we also plot the
bound on negative curvature calculated by Freivogel et
al. \cite{Freivogel:2005vv} (which are similar to those of Vilenkin
and Winitzki \cite{Vilenkin:1996ar} and Garriga et
al. \cite{Garriga:1998px}) by demanding that structures at least as
large as a small sized galaxy form. Our plot allows for somewhat more
curvature than is allowed by these 
methods. Setting a structure formation limit based on smaller galactic
masses brings the curvature limit closer to ours. Ultimately our rough
reproduction of the anthropic cutoff is unsurprising as our
main entropy source, star formation, cuts off along with structure
formation. However, our actual prediction for curvature is not
against a cutoff for structure formation as would be the case for a
simple bound. The causal entropic weighting provides additional
rewards for smaller curvatures in the form of increased star formation
and entropy production.

\section{\label{sec:Sec4} Conclusions}
Anthropic constraints on observable parameters are interesting when
considering implications for the multiverse. The CEP has appealing
advantages over previous anthropic attempts. We find that the CEP
places upper limits on the amount of curvature that is observable and
while not ruling out our universe, the CEP finds larger curvature
preferable to our measured value. We also find an intriguing feature
in the probability space for positive curvature of an elongated tail
stretching into regions of large curvature. Our results for negatively
curved universes are broadly consistent with previous anthropic bounds
on curvature but less constraining due to a more lenient tolerance for
the minimum mass of structure allowed.  Still, like the previous work we
find that anthropic considerations seem to offer cosmology
considerable relief from the flatness problem.

\begin{acknowledgements} We thank
Lloyd Knox, Damien Martin, and especially James Cline for
  very helpful discussions. We also thank Tony Tyson for computing
  resources as well as Perry Gee and Jim Bosch for technical computing support. This work was
  supported by DOE grant DE-FG03-91ER40674.
\end{acknowledgements}

\bibliography{CEPC.bib}
\end{document}